%% file: jea2014.tex
\documentclass{article}

\usepackage[english]{babel}
\usepackage[T1]{fontenc}
\usepackage{lmodern}
\usepackage{xspace}
\usepackage{fullpage}
\usepackage{epsfig}
\usepackage{graphics}
\usepackage{graphicx}
\usepackage{color}
\usepackage{amsthm}
\usepackage{amsmath,amssymb}
\usepackage{latexsym}
\usepackage{array}
\usepackage{slashbox}
\usepackage{multirow}
\usepackage{url}
\usepackage[ruled,vlined]{algorithm2e}
\usepackage[autolanguage,np]{numprint}

\newcommand{\inC}{\ensuremath{\textsf{\upshape in}[C]}\xspace}
\newcommand{\inCp}{\ensuremath{\textsf{\upshape in}^{\prime}[C]}\xspace}
\newcommand{\totC}{\ensuremath{\textsf{\upshape tot}[C]}\xspace}
\newcommand{\wC}{\ensuremath{\textsf{\upshape s}[C]}\xspace}
\newcommand{\dnc}{\ensuremath{d_w(i,C)}\xspace}
\newcommand{\dncp}{\ensuremath{d^{\prime}_w(i,C)}\xspace}

\newcolumntype{M}[1]{>{\raggedright}m{#1}}

\renewcommand{\algorithmcfname}{ALGORITHM}
\SetAlFnt{\small}
\SetAlCapFnt{\small}
\SetAlCapNameFnt{\small}
\SetAlCapHSkip{0pt}
\IncMargin{-\parindent}

\newcommand{\keywords}[1]{\bigskip \textbf{Key words:} \itshape #1 \normalfont}

\title{A Generalized and Adaptive Method for Community Detection\thanks{~This work is partially supported by the \textsl{CODDDE ANR-13-CORD-0017-01} project of the French National Research Agency and by the \textsl{CLEAR} common project between Thales and the Paris 06 university.}}

\author{Romain Campigotto\thanks{Sorbonne Universit\'{e}s, Univ Paris 06, UMR 7606, LIP6, F-75005, Paris, France}~$^\text{,}$\thanks{CNRS, UMR 7606, LIP6, F-75005, Paris, France}\\ \url{romain.campigotto@lip6.fr}
  \and
  Patricia Conde C\'{e}spedes\thanks{Sorbonne Universit\'{e}s, Univ Paris 06, EA 3124, LSTA, F-75005, Paris, France}\\ \url{patricia.conde_cespedes@upmc.fr}
  \and
  Jean-Loup Guillaume\footnotemark[2]~$^\text{,}$\footnotemark[3]\\ \url{jean-loup.guillaume@lip6.fr}
}

\date{}

\begin{document}

\maketitle

\begin{abstract}
  Complex networks represent interactions between entities. They appear in various contexts such as sociology, biology, etc., and they generally contain highly connected subgroups called \emph{communities}. Community detection is a well-studied problem and most of the algorithms aim to maximize the \textsl{Newman-Girvan} modularity function, the most popular being the Louvain method (it is well-suited on very large graphs). However, the classical modularity has many drawbacks: we can find partitions of high quality in graphs without community structure, e.g., on random graphs; it promotes large communities. Then, we have adapted the Louvain method to other quality functions.

  In this paper, we describe a generic version of the Louvain method. In particular, we give a sufficient condition to plug a quality function into it. We also show that global performance of this new version is similar to the classical Louvain algorithm, that promotes it to the best rank of the community detection algorithms.
  
\end{abstract}

\input{intro.tex}

\input{algo.tex}

\input{criteria.tex}

\input{gen.tex}

\input{exp.tex}

\input{conclusion.tex}

\section*{Acknowledgements}

The authors would like to thank No\'{e} Gaumont for his improvements on the generic Louvain method software and Thales for their funding of this research.


\bibliographystyle{abbrv}
\bibliography{references}

\end{document}

%% file: intro.tex
\section{Introduction}
\label{sect:intro}

Complex networks are very common in different areas such as sociology (e.g., friendship or collaboration networks), computer science (e.g., networks of computers), biology (e.g., gene regulatory networks or neural networks). These networks are often modeled using the graph formalism which represents the interactions between a set of entities. For example, a social network can be represented by a graph whose nodes (or \emph{vertices}) are individuals and links (or \emph{edges}) represent a kind of social relationship.

Most complex networks exhibit a community structure, i.e. are composed of groups of highly connected nodes (e.g., groups of friends or collaborators)~\cite{girvan2002community,SKP-2012}. The automatic detection of such communities has attracted much attention in recent years and many community detection algorithms have been proposed (see~\cite{fortunato2010community} for a survey). In most studies, the detection of communities aims at finding a partition of the nodes of the graph, i.e. any given node belongs to exactly one community, and communities are, therefore, groups of highly connected nodes and loosely connected to the rest of the graph. A classical way to assess the quality of a partition in communities is to maximize the quality function known as \emph{modularity}~\cite{newman2004finding}, which measures the difference between the effective and the expected internal link density in communities. However, due to the intrinsic \textbf{NP}-hardness of linear programming with binary variables subject to the linear constraints of a partition (proved in~\cite{GW-1989} and after in~\cite{Waka-1998} for the acyclic subgraph problem), modularity maximization is an \textbf{NP}-hard problem~\cite{brandes2007finding} and most algorithms use heuristics. Among the formers, one of the most efficient (both in speed and quality) is the \emph{Louvain method}~\cite{ABGL-2011,BGLL-2008}, which is an hierarchical local greedy technique to maximize modularity.

While modularity (and its maximization) has gained a lot of attention in the complex networks field since it was proposed, many other quality functions have been defined in close or very different contexts, and the first notable proposition, derived from the works of Condorcet about voting consensus~\cite{CON1785}, was made by \textsl{C. T. Zahn}~\cite{ZAH64} in the sixties.

\medskip

Given that other functions have not been studied deeply in the complex network field, we propose here a generic version of the Louvain algorithm that can be used on a large variety of criteria. To achieve this, we show that the Louvain method core is mainly independent of the underlying quality function and, therefore, that it can be generalized to many other quality functions. We also give a sufficient condition for a quality function which ensures that it can be \emph{efficiently} plugged into the Louvain framework.

\bigskip

Our paper is organized as follows: in Sect.~\ref{sect:algo} we present the Louvain method and we point out the portions of the algorithm which are directly used for modularity maximization. In Sect.~\ref{sect:criteria} we present a large variety of criteria to evaluate a partition and we show two properties of such criteria that can can be used to decide whether a criterion can be plugged into Louvain or not. In Sect.~\ref{sect:gen} we give the generic version of the Louvain algorithm and we instantiate it for another criterion. We also give a sufficient condition for a quality function to be efficiently pluggable into Louvain and give an example of a criterion which does not meet this condition but can be efficiently plugged (since the condition is not necessary). Finally, we show in Sect.~\ref{sect:exp} that the generic version of the Louvain method is very efficient for most criteria and that it can successfully replace the classical Louvain algorithm.

%% file: algo.tex
\section{The Louvain method}
\label{sect:algo}

\begin{figure}[t]
      \centering
      \includegraphics[width=0.6\paperwidth]{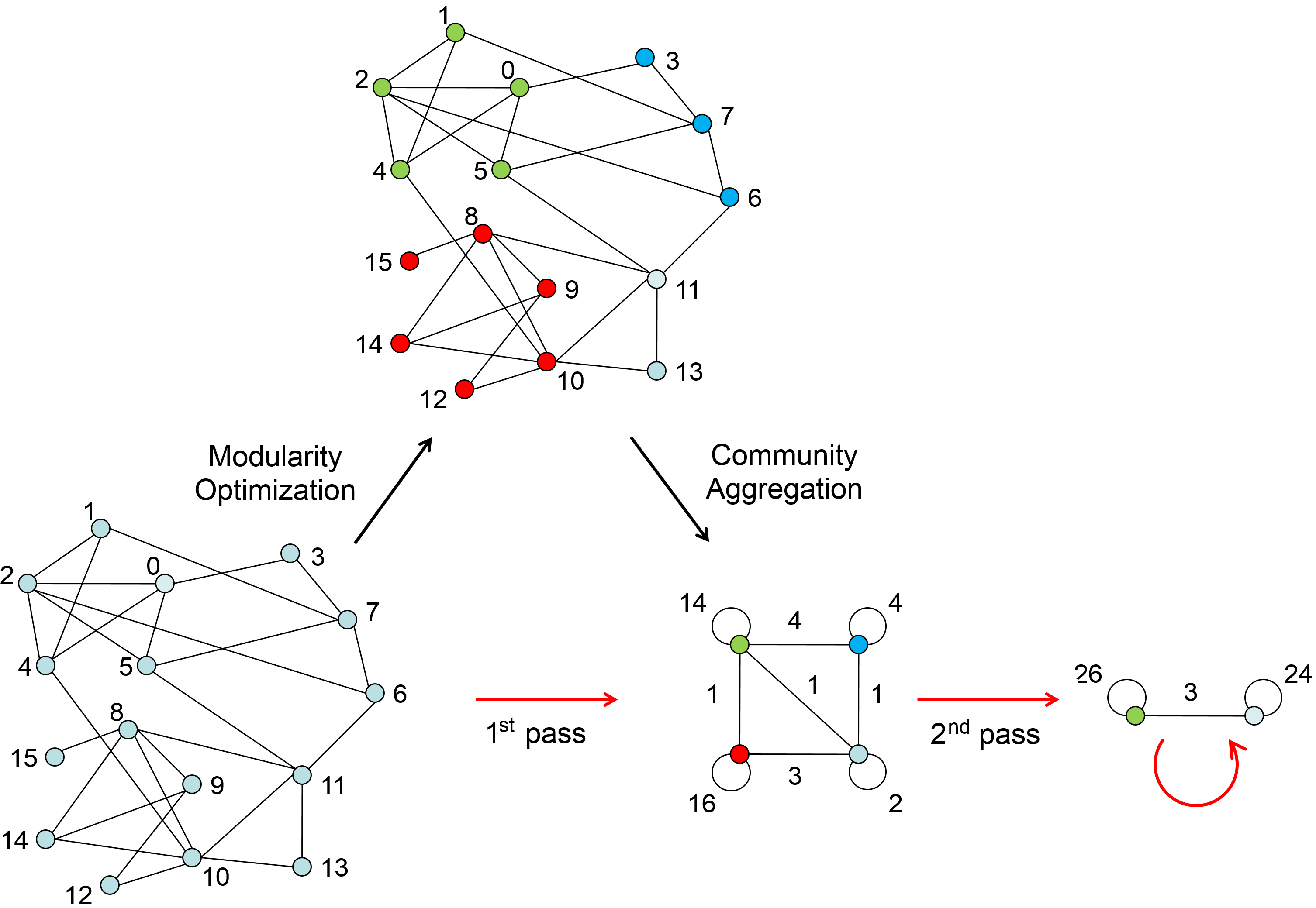}
      \caption{Louvain algorithm overview (Fig.~1 in~\cite{BGLL-2008})}
      \label{fig:ex_levels}
    \end{figure}

The Louvain algorithm has first been introduced to find partitions of high \textsl{Newman-Girvan} modularity, which is the most widely used criterion to evaluate a partition of a complex network. The algorithm uses a local greedy optimization to find a local maxima of modularity, which is further enhanced with a hierarchical refinement. The local greedy optimization is described in Algorithm~\ref{algo:louvainonepass}: it consists in moving nodes one by one in one of their neighboring community so as to obtain the maximal increase of modularity. Nodes can be moved several times and this procedure stops only when a local maxima is obtained, i.e. when no individual move can increase the modularity. The hierarchical refinement described in Algorithm~\ref{algo:louvain} consists in building a meta-graph whose nodes are the communities found in the previous step and links represent the sum of connections between the communities (see Fig.~\ref{fig:ex_levels} for an example). The Louvain algorithm is therefore an iteration of two sub-procedures: local optimization and meta-graph construction. It stops only when no improvement can be obtained by any of the two operations.

\medskip

Several improvements of the Louvain algorithm have been proposed in the literature. These improvements are directed either towards an improvement of the time efficiency mainly by using less severe stopping criteria, or towards an improvement of the quality of the result by using refinements, such as the ones proposed in~\cite{dMFFP-2011,GH-2013}, or an improvement of the computation time by parallelization~\cite{BS-2013}. However, we will mainly focus on the simple version of the algorithm and most refinements can be used on the generic version of Louvain.

\begin{algorithm}[t]
  \caption{One pass algorithm}
  \label{algo:louvainonepass}
  
  \DontPrintSemicolon
  
  \SetKwInput{Require}{Require}
  \SetKwInput{Ensure}{Ensure}
  \SetKwInput{Data}{Local}
  
  \Require{$G=(V,E,w)$ a weighted graph}
  \Ensure{a partition $\mathcal{P}$ of $V$}
  
  \BlankLine
  
  \Data{$increase \leftarrow \texttt{true}$}
  \Data{$\mathcal{P}$ the current partition of $V$}
  
  \BlankLine
  
  \Begin{
      
      \ForAll{\textup{nodes} $i$}{
        $\mathcal{P}[i] \leftarrow \{i\}$\;
        INIT$(i)$\;
      }

      \While{$increase$}{
        $increase \leftarrow \texttt{false}$\;
        \ForAll{\textup{nodes} $i$}{
          $C_{\text{old}} \leftarrow \mathcal{P}[i]$\;
          REMOVE$(i,C_{\textup{old}})$\;
          $\mathcal{C} \leftarrow \{\mathcal{P}[j] \mid (i,j)\in E\} \cup \{C_{\text{old}}\}$\;
          $C_{\text{new}} \leftarrow \arg\max_{C \in \mathcal{C}}\{$GAIN$(i,C)\}$\;
          INSERT$(i,C_{\text{new}})$\;
          \If{$C_{\textup{old}} \neq C_{\textup{new}}$} {
            $increase \leftarrow \texttt{true}$\;
          }
        }
      }
      
    }
    
  \end{algorithm}
  
  \begin{algorithm}[t]
    \caption{Louvain algorithm}
    \label{algo:louvain}
    
    \DontPrintSemicolon
    
    \SetKwInput{Require}{Require}
    \SetKwInput{Ensure}{Ensure}
    
    \Require{$G=(V,E,w)$ a weighted graph}
    \Ensure{a partition $\mathcal{P}$ of $V$}
    
    \BlankLine
    
    \Begin{
        \Repeat{\upshape no improvement is possible}{
          $\mathcal{P} \leftarrow $ONEPASS$(G)$\;
          $G \leftarrow $PARTITION-TO-GRAPH$(\mathcal{P},G)$\;
        }
      }
      
    \end{algorithm}

    \medskip
    
    The Louvain algorithm presented in Algorithms~\ref{algo:louvainonepass} and~\ref{algo:louvain} contains several sub-functions that must be precisely defined and are strongly related to the \textsl{Newman-Girvan} modularity function to be optimized. Basically, one must define functions to remove (resp. insert) a node from (resp. in) a community and update all the information needed by the algorithm, and a function to compute the gain (or loss) in the value of modularity when a node is moved in a given community.
    
    For the classical Louvain method, we give the four functions in Algorithm~\ref{algo:louvainfunctions}. These functions have been originally described in~\cite{BGLL-2008} and later in~\cite{ABGL-2011}. We precise here that
\begin{eqnarray}
  \label{eqn:dnc}
  \dnc & = & \sum_{j \in C} w_{ij} \enspace ,
\end{eqnarray}
with $w_{ij}$ the link weight between nodes $i$ and $j$. They are based on the use of two vectors, \textsf{in} and \textsf{tot}, which indicate, for each community, the number of intern and total links. These vectors are updated at each insertion or removal and are used to compute the modularity gain for each move. Note that the $\textsf{in}$ vector is not useful since the gain in modularity when a node is moved to community $C$ can be written in a more compact form as follows
    \begin{equation}
      \label{eqn:gain_mod}
      \dnc - \frac{d_i \cdot \totC}{2m} \enspace ,
    \end{equation}
    with $d_i = \sum_{j \mid (i,j) \in E} w_{ij}$ the \emph{weighted degree} of node $i$ and $2m = \sum_{i \in V} d_i$; but it is necessary to compute the complete quality of the graph partition.
    
    \begin{algorithm}[t]

      \caption{Functions for the \textsl{Newman-Girvan} modularity}
      \label{algo:louvainfunctions}
      
      \DontPrintSemicolon
      
      INIT$(i)$:\;
      \hspace*{1mm}\Begin{
          $\textsf{in}[i] \leftarrow w_{ii}$\;
          $\textsf{tot}[i] \leftarrow d_i$\;
        }

        \BlankLine

      REMOVE$(i,C)$:\;
      \hspace*{1mm}\Begin{
          $\inC \leftarrow \inC - 2 \cdot \dnc - w_{ii}$\;
          $\totC \leftarrow \totC - d_i$\;
          $\mathcal{P}[i] \leftarrow \emptyset$\;
        }

        \BlankLine
        
        INSERT$(i,C)$:\;
        \hspace*{1mm}\Begin{
            $\inC \leftarrow \inC + 2 \cdot \dnc + w_{ii}$\;
            $\totC \leftarrow \totC + d_i$\;
            $\mathcal{P}[i] \leftarrow C$\;
          }

          \BlankLine
          
          GAIN$(i,C)$:\;
          \hspace*{1mm}\Begin{
              return~$\displaystyle \left[\frac{\inC + 2 \cdot \dnc}{2m} - \left(\frac{\totC + d_i}{2m}\right)^2 \right] - \left[\frac{\inC}{2m} - \left(\frac{\totC}{2m}\right)^2 - \left(\frac{d_i}{2m}\right)^2 \right]$
            }
            
          \end{algorithm}

          These four functions can be designed for other criteria, that we will show in Sect.~\ref{sect:gen}. Before that, we will present in Sect.~\ref{sect:criteria} some other criteria for graph clustering.

%% file: criteria.tex
\section{Some Criteria for Graph Clustering}
\label{sect:criteria}

Graph clustering criteria depend strongly on the meaning given to the notion of \emph{community}. In the last few years, various modularization criteria have been defined in different fields, each one having its own definition of \emph{community}. Although different, all definitions have something in common: \textit{dense connections within communities and only sparse connections between communities}.

In this section, we present different modularization criteria in Relational coding\footnote{For more details about Relational Analysis theory, see~\cite{MAR84}.}; this notation will help us to compare those criteria on the same basis.

\medskip

Since a graph represents relations between objects belonging to the same set, therefore, a non-oriented and non-weighted graph $G=(V,E)$, with $n=|V|$ nodes and $m=|E|$ links, is a binary symmetric relation on its set of nodes $V$ represented by its adjacency matrix $\textbf{A}$, whose general term is defined as
\begin{equation}
  \label{eqAdjacenceatrix}
  a_{ij} =
  \begin{cases}
    1 & \text{if there exists an link between $i$ and $j$ $\forall (i,j) \in V \times V$,}\\
    0 & \text{otherwise.}
  \end{cases}
\end{equation}

In the case of a weighted graph $G=(V,E,w)$, the adjacency matrix $\textbf{A}$ will be denoted $\textbf{W}$. The general term of this matrix, $w_{ij}$, represents the weight of the edge linking nodes $i$ and $j$.

Partitioning a graph implies to define an equivalence relation on the set of nodes $V$, that means a symmetric, reflexive and transitive relation. Mathematically, an equivalence relation is represented by a square matrix $\textbf{X}$ whose entries are defined as
\begin{equation}
  \label{eqinconnueX}
  x_{ij} =
  \begin{cases}
    1 & \text{if node $i$ and node $j$ are in the same community $\forall (i,j) \in V \times V$,}\\
    0 & \text{otherwise.}
  \end{cases}
\end{equation}

Modularizing a graph implies to define $\textbf{X}$ as close as possible to $\textbf{A}$. A modularity criterion is a function which measures either a similarity or a distance between $\textbf{A}$ and $\textbf{X}$. Therefore, the problem of modularization can be written as a function to optimize (\emph{minimize} if the function represents a distance or \emph{maximize} if the function represents a similarity). So, we have
\begin{equation}
  \max_X~\text{ or }~\min_X F(X) \enspace ,
\end{equation}
subject to the constraints of an equivalence relation:
\begin{eqnarray}
  \label{eqContraintesRel Equiv}
  x_{ij}\in\left\{0,1\right\} & & \text{Binary}\nonumber\\
  x_{ii}=1 & \forall i & \text{Reflexivity}\nonumber\\
  x_{ij}-x_{ji}=0 & \forall(i,j) & \text{Symmetry}\nonumber\\
  x_{ij}+x_{jk}-x_{ik}\le 1 & \forall(i,j,k) & \text{Transitivity}\nonumber
\end{eqnarray}

We define as well $\bar{\textbf{X}}$ and $\bar{\textbf{A}}$ as the inverse relation of $\textbf{X}$ and $\textbf{A}$ respectively. Their entries are defined as $\bar{x}_{ij}=1-x_{ij}$ and $\bar{a}_{ij}=1-a_{ij}$ respectively. In the case of a weighted graph, the general term of the matrix $\bar{\textbf{W}}$ is calculated as follows: $\bar{w}_{ij}= \mathcal{W}-w_{ij}$, where $\mathcal{W} = \max_{(i,j) \in E} \{w_{ij}\}$.

\bigskip

We distinguish the criteria according to the properties they verify. We consider two properties: \emph{linearity} and \emph{separability}. The relational codings of these properties are shown in Table~\ref{tabPropriBonCriter}.

\begin{table}[!ht]
  \caption{Properties verified by modularity criteria$^a$}  
  
  \begin{center}
    \begin{tabular}{|r|l|}
      \hline
      &\\
      \textbf{The criterion has the property} & \textbf{If it can be written as} \\&\\\hline
      &\\
      Linearity & $\displaystyle{F(X)~=~\sum_{i,j \in V} \phi(a_{ij})x_{ij}+K}$ \\\hline
      &\\
      Separability & $\displaystyle{F(X)~=~\sum_{i,j \in V} \phi( a_{ij})\psi(x_{ij})+K}$ \\\hline
    \end{tabular}

    \medskip
    
    $^a${\footnotesize We use $K$ to denote any constant depending only on the original data.}
  \end{center}

  \label{tabPropriBonCriter}
\end{table}

According to Table~\ref{tabPropriBonCriter}, the property of linearity entails that the criterion is a linear function of $\textbf{X}$. The function $\phi(a_{ij})$ depends only on the original data (i.e. the adjacency matrix). $\psi(x_{ij})$ is a function of the unknown variable $x_{ij}$. The property of separability implies that the criterion can be written as a scalar product of two vectors, the first one depending only upon the original data and the second one depending upon the unknown variable. Consequently, \textit{every linear criterion is separable}.

\subsection{Linear Criteria (and Separable)}

The Relational coding of all linear criteria is given in Table~\ref{tabCriterLinSep}. For each criterion, the optimal partition is found without fixing in advance the optimal number of \emph{clusters} (or \emph{communities}) $\kappa$. Just for comparison, we added the \textsl{Newman-Girvan} modularity.

\begin{table}[t]
  \centering
  \begin{tabular}{|r|l|}
    \hline
    &\\
    \textbf{Criterion} & \textbf{Relational notation} \\&\\\hline
    &\\
    \textsl{Newman-Girvan} (2004) & $\displaystyle F_{\textsc{ng}}(X)~=~\sum_{i,j \in V} \left(a_{ij}-\frac{d_id_j}{2m}\right)x_{ij}$ \\\hline
    &\\
    \textsl{Zahn-Condorcet} (1785, 1964) & $\displaystyle F_{\textsc{zc}}(X)~=~\sum_{i,j \in V} a_{ij}x_{ij} + \sum_{i,j \in V} \bar{a}_{ij}\bar{x}_{ij}$ \\\hline
    &\\
    \textsl{Owsi\'nski-Zadro\.zny} (1986) & $\displaystyle F_{\textsc{oz}}(X)~=~(1-\alpha) \sum_{i,j \in V} a_{ij}x_{ij} + \alpha \sum_{i,j \in V} \bar{a}_{ij}\bar{x}_{ij}$ \newline{with $0<\alpha<1$} \\\hline
    &\\
    \textsl{Marcotorchino} (1991) & $\displaystyle F_{w\textsc{c}}(X)~=~\sum_{i,j \in V} \hat{a}_{ij}x_{ij} + \sum_{i,j \in V} \bar{\hat{a}}_{ij}\bar{x}_{ij}$ \\\hline
    &\\
    Balanced Modularity (2013) & $\displaystyle{F_{\textsc{bm}}(X)~=~\sum_{i,j \in V} \left(a_{ij}-\frac{d_id_j}{2m}\right)x_{ij} + \sum_{i,j \in V} \left(\bar{a}_{ij}-\frac{(n-d_i)(n-d_j)}{n^2-2m}\right)\bar{x}_{ij}}$ \\\hline
    &\\
    Deviation to Indetermination (2013) & $\displaystyle{F_{\textsc{di}}(X)~=~\sum_{i,j \in V} \left(a_{ij}-\frac{d_i}{n}-\frac{d_j}{n}+\frac{2m}{n^2}\right)x_{ij}}$ \\\hline
    &\\
    Deviation to Uniformity (2013) & $\displaystyle{F_{\textsc{du}}(X)~=~\sum_{i,j \in V} \left(a_{ij}-\frac{2m}{n^2} \right)x_{ij}}$ \\\hline
  \end{tabular}
  \caption{Linear modularity functions in Relational notation}  
  \label{tabCriterLinSep}
\end{table}

\begin{enumerate}

\item \textbf{The \textsl{Newman-Girvan} criterion (2004)} (see~\cite{newman2004finding}): also known as \emph{modularity}, its definition involves a comparison of the number of within-cluster links in a real network and the expected number of such links in a random graph (without regard to community structure).

  \medskip

\item \textbf{The \textsl{Zahn-Condorcet} criterion (1785, 1964)}: \textsl{C. T. Zahn} (see~\cite{ZAH64}) was the first author who studied the problem of finding an equivalence relation $\textbf{X}$, which best ``approximates'' a given symmetric relation $\textbf{A}$ in the sense of minimizing the distance of the symmetric difference. However, the criterion defined by \textsl{Zahn} is a variant of the relational \textsl{Condorcet}'s criterion (see~\cite{CON1785}), introduced in his works on Voting Consensus and whose relational notation was first given by~\cite{MAM79}.
  
  \medskip
  
\item \textbf{The \textsl{Owsi\'nski-Zadro\.zny} criterion (1986)} (see~\cite{OWZ86}) is a generalization of \textsl{Zahn-Condorcet}'s function. It is more flexible because it contains a parameter $\alpha$ which allows the user, according to the context, to define the minimal percentage $\alpha$ of required within-cluster links in each community.

  \medskip

\item \textbf{The \textsl{Marcotorchino} criterion (1991)} (or the A-weighted \textsl{Condorcet} criterion): introduced in~\cite{MAR91a} (see also~\cite{HUQ91}), it is a derivation of the \textsl{Zahn-Condorcet}'s criterion. Instead of considering the adjacency matrix $\textbf{A}$, we consider the weighted A-matrix\footnote{This matrix plays an important role in Factorial Correspondence Analysis (FCA).} $\hat{\textbf{A}}$ and its complementary $\bar{\hat{\textbf{A}}}$, which are defined as
  \begin{equation}
    \label{eqCpondHat}
    \hat{a}_{ij}~=~\frac{2 a_{ij}}{d_i+d_j} \quad \text{and} \quad \bar{\hat{a}}_{ij}~=~\frac{\hat{a}_{ii}+\hat{a}_{jj}}{2}-\hat{a}_{ij} \enspace ,
  \end{equation}
  where $d_i=\sum_{j \in V} a_{ij}$ is the degree of node $i$. If the graph does not contain loops, the optimization of this criterion groups all nodes together. Therefore, before adapting this modularity function, we must add self-loops to all nodes of the graph. Moreover, the use of this criterion is limited to non-weighted graphs.

\medskip

\item {\textbf{The Balanced Modularity criterion (2013)}}: introduced first in~\cite{COM13}, it is a balanced version of the \textsl{Newman-Girvan} modularity. In fact, it was constructed by adding to the \textsl{Newman-Girvan} modularity a term taking into account the absence of links $\bar{\textbf{A}}$. 
  
  \medskip

\item \textbf{The Deviation to Indetermination criterion (2013)}: analogously to \textsl{Newman-Girvan} function, which maximizes the deviation to the independence structure, this index maximizes the deviation to the indetermination structure (see~\cite{JAV82},~\cite{AHM07} and~\cite{MAR13}).

  \medskip

\item \textbf{The Deviation to Uniformity criterion (2013)}: proposed in~\cite{CON13}, this criterion maximizes the deviation to the uniformity structure. It compares the number of within-cluster edges in the real graph and the expected number of such edges in a random graph where edges are uniformly distributed, thus all the nodes have the same degree equal to the average degree of the graph. We can compare such a network to grid or a lattice graph with no community structure.

\end{enumerate}

\bigskip

We can see from Table~\ref{tabCriterLinSep} that all the criteria can be turned into the general form
\begin{equation}
  \label{eqGeneralLinSepCr}
  F(X)~=~\sum_{i,j \in V}\left( \phi(a_{ij})x_{ij}+\bar{\phi}(\bar{a}_{ij})\bar{x}_{ij}\right) \enspace ,
\end{equation}
where $\phi$ and $\bar{\phi}$ are functions of the original data (in some cases, we have $\bar{\phi}(.)~=~0$: for instance, this is the case of the Deviation to Indetermination index). From (\ref{eqGeneralLinSepCr}) and the definition of $\bar{\textbf{X}}$, we can infer that every linear criterion can be written as
\begin{equation}
  \label{eqGeneralLinSepCr2}
  F(X)~=~\sum_{i,j \in V}\left( \phi(a_{ij})-\bar{\phi}(\bar{a}_{ij})\right) x_{ij}+K \enspace .
\end{equation}

As the unknown variable $x_{ij}$ is null if $i$ and $j$ are not in the same community, (\ref{eqGeneralLinSepCr2}) allows to evaluate the function by summing up terms depending only on intra-cluster properties. Furthermore, this expression demonstrates that every linear criterion is separable.

\subsection{Non-Linear Criteria}

These criteria are non-linear functions of the unknown variable $\textbf{X}$; their expressions are shown in Table~\ref{tabCriterSepNonLin}. We can deduce from this table that all criteria are separable, except the \textsl{Shi-Malik} function (also known as \emph{normalized cuts} criterion).

\begin{table}[t]
  \centering
  \begin{tabular}{|r|l|}
    \hline
    &\\
    \textbf{Criterion} & \textbf{Relational notation} \\&\\\hline
    &\\
    Profile Difference (1976) & $\displaystyle{F_{\textsc{pd}}(X)~=~2 \sum_{i,j \in V} \hat{a}_{ij}\hat{x}_{ij}-\kappa - \sum_{i,j \in V} \hat{a}_{ij}^2}$ \\\hline
    &\\
    \textsl{Mancoridis-Gansner} (1998) & $\displaystyle F_{\textsc{mg}}(X)~=~\frac{1}{\kappa}\sum_{i,j \in V} \frac{a_{ij}x_{ij}}{|C_{[i]}| \cdot |C_{[j]}|}+\frac{1}{\kappa(\kappa-1)} \sum_{i,j \in V} \frac{\bar{a}_{ij}\bar{x}_{ij}}{|C_{[i]}| \cdot |C_{[j]}|}\quad$ \newline with $\kappa >1$ \\\hline
    &\\
    \textsl{Michalski-Decaestecker} (1983) & $\displaystyle F_{\textsc{md}}(X)~=~\sum_{i,j \in V} a_{ij}\hat{x}_{ij} + \sum_{i,j \in V} \bar{a}_{ij}\bar{\hat{x}}_{ij}$ \\\hline
    &\\
    \textsl{Shi-Malik} (2000) & $\displaystyle{F_{\textsc{sm}}(X)~=~\sum_{i,j \in V} \frac{a_{ij}\bar{x}_{ij}}{\sum_{k \in V} d_k x_{ik}}}$ \\\hline
    &\\
    \textsl{Michalski-Goldberg} (2012) & $\displaystyle{F_{\textsc{g}}(X)~=~\sum_{i,j \in V} a_{ij}\hat{x}_{ij}}$ \\\hline
  \end{tabular}
  \caption{Non-linear criteria}    
  \label{tabCriterSepNonLin}
\end{table}

\begin{enumerate}

\item \textbf{The Profile Difference criterion (1976)}: introduced as a contingency measure of the distance between two partitions (see~\cite{CAP76} and~\cite{BED92}), it minimizes a least square function: the \emph{squared Frobenius norm} of the difference between the matrices $\hat{\textbf{A}}$ and the weighted X-matrix $\hat{\textbf{X}}$, defined as
  \begin{equation}
    \label{eqn:xhat}
    \hat{x}_{ij}~=~\frac{2 x_{ij}}{|C_{[i]}|+|C_{[j]}|} \quad \text{and} \quad \bar{\hat{x}}_{ij}~=~\frac{\hat{x}_{ii}+\hat{x}_{jj}}{2}-\hat{x}_{ij} \enspace ,
  \end{equation}
  with $|C_{[i]}|$ the size of the community containing the node $i$. The original expression of this criterion is given in relational terms by
  \begin{eqnarray}
    \label{eqn:ProfileDifference}
    F_{\textsc{pd}}(X) & = & ||\hat{\textbf{A}}-\hat{\textbf{X}}||^2~=~\sum_{i,j \in V} \left(\hat{a}_{ij}-\hat{x}_{ij}\right)^2 \enspace .
  \end{eqnarray}
  Since $\textbf{X}$ represents an equivalence relation, we have
  \begin{eqnarray}
    \label{eqn:xhatij}
    \hat{x}_{ij} & = & \frac{x_{ij}}{|C_{[i]}|}~=~\frac{x_{ij}}{|C_{[j]}|} \enspace .
  \end{eqnarray}
  Therefore, minimizing this criterion is equivalent to maximize its expression given in Table~\ref{tabCriterSepNonLin}, where $\kappa$ represents the number of communities (or clusters) of the optimal partition.

  \medskip

\item \textbf{The \textsl{Mancoridis-Gansner} criterion (1998)}: originally introduced to recover the modular structure of a software from its source code in~\cite{MAN98}, this criterion aims simultaneously at maximizing intra-connectivity within clusters and minimizing inter-connectivity between clusters. The intra-connectivity of a cluster is calculated as the fraction of the maximum number of intra-cluster links. The inter-connectivity between two clusters is calculated as the fraction of the maximum number of inter-clusters links. Both measures, the intra-connectivity and the inter-connectivity, are normalized by the number of clusters $\kappa$ and the number of distinct pair of clusters $\frac{\kappa(\kappa-1)}{2}$.

  \medskip

\item \textbf{The \textsl{Michalski-Decaestecker} criterion (1983)} (or the X-weighted \textsl{Condorcet} criterion): introduced by~\cite{MIS83} in Learning theory, its relational expression (given in~\cite{DEC92}) is a derivate of \textsl{Zahn-Condorcet} criterion. This criterion is built up by replacing the unknown variables $x_{ij}$ with the weighted variables $\hat{x}_{ij}$. Unlike the criteria previously presented, for this one, it is necessary to fixe the number of clusters $\kappa$ in advance, otherwise, the optimal solution is trivial: each node is isolated and, therefore, there are $n$ clusters ($\kappa=n$).

\medskip

\item \textbf{The \textsl{Shi-Malik} criterion (2000)} (or \emph{normalized cuts}: see~\cite{SHM00}): this function is of type \emph{min-cut}, a type of criterion which minimizes the number of \emph{inter-cluster links} (better known as \emph{cut edges}) or a function of this number. The \textsl{Shi-Malik} criterion minimizes the \emph{cut edges} as a fraction of the total link connections to all the nodes in the graph. If the number of clusters is not fixed in advance, minimizing the \emph{cut edges} turns out to group all the nodes together. Hence, without fixing in advance $\kappa$, the optimal solution is trivial since all the nodes are clustered together.

\medskip

\item \textbf{The \textsl{Michalski-Goldberg} criterion (2012)}: (see ~\cite{Gold-1984}): this function maximizes the density of edges in each community. Thus, for each cluster, the density of \textsl{Goldberg} is the number of within-cluster edges divided by the size of the cluster (it corresponds to the first term of the \textsl{Michalski-Decaestecker} criterion previously defined). It has been already used for graph partitionning in~\cite{DBM-2012}.

\end{enumerate}

\bigskip

In the next section, we will show that all the linear criteria can be plugged easily into the Louvain method, i.e. we can write the three specific functions to insert (resp. remove) a node into (resp. from) a community and compute \textbf{efficiently} (i.e. just by re-computing locally the quality function for the community considered) the gain obtained when a node is moved in a given community.

%% file: gen.tex
\section{A Generic Version of Louvain Algorithm}
\label{sect:gen}

Algorithms~\ref{algo:louvainonepass} and~\ref{algo:louvain} page~\pageref{algo:louvainonepass} already give the generic version of Louvain algorithm. Indeed, this local optimization heuristic can be easily adapted to other quality functions than modularity. In fact, only the functions INIT, REMOVE, INSERT and GAIN described in Algorithm~\ref{algo:louvainfunctions} are specifically designed for the \textsl{Newman-Girvan} modularity; but we can adapt these functions for other criteria: for example, Algorithm~\ref{algo:ZCfunctions} explicits the functions for the \textsl{Zahn-Condorcet} quality function, where $s_i$ denotes the size of (meta-)node $i$ ($\forall i \in V$, $s_i=1$ for the first level of Louvain algorithm) and $\wC$ denotes the size of community $C$ ($\wC = \sum_{i \in C} s_i$).\footnote{As Louvain groups communities into meta-nodes, we have to save the number of nodes contained into a meta-node. This is needed for several criteria, like \textsl{Zahn-Condorcet} (but this is not the case for the \textsl{Newman-Girvan} modularity).}

\medskip

\begin{algorithm}[t]
  
  \caption{Functions for the \textsl{Zahn-Condorcet} quality function}
  \label{algo:ZCfunctions}
  
  \DontPrintSemicolon
  
  INIT$(i)$:\;
  \hspace*{1mm}\Begin{
      $\textsf{in}[i] \leftarrow w_{ii}$\;
      $\textsf{s}[i] \leftarrow s_i$\;
    }
    
    \BlankLine
    
    REMOVE$(i,C)$:\;
    \hspace*{1mm}\Begin{
        $\inC \leftarrow \inC - 2 \cdot \dnc - w_{ii}$\;
        $\wC \leftarrow \wC - s_i$\;
        $\mathcal{P}[i] \leftarrow \emptyset$\;
      }
      
      \BlankLine
      
      INSERT$(i,C)$:\;
      \hspace*{1mm}\Begin{
          $\inC \leftarrow \inC + 2 \cdot \dnc + w_{ii}$\;
          $\wC \leftarrow \wC + s_i$\;
          $\mathcal{P}[i] \leftarrow C$\;
        }
        
        \BlankLine
        
        GAIN$(i,C)$:\;
        \hspace*{1mm}\Begin{
            return~$\displaystyle 2 \cdot \dnc - \mathcal{W} \cdot s_i \cdot \wC$
          }
          
        \end{algorithm}

To obtain the GAIN function described in Algorithm~\ref{algo:ZCfunctions}, we have to transform the Relational notation of the criterion given in Table~\ref{tabCriterLinSep} page~\pageref{tabCriterLinSep}.

We recall here (for the weighted case) the Relational notation of the \textsl{Zahn-Condorcet} criterion:
\begin{eqnarray}
  \label{eqn:ZCmatrix}
  F_{\textsc{zc}}(X) & = & \sum_{i,j \in V} w_{ij}x_{ij} + \sum_{i,j \in V} \bar{w}_{ij}\bar{x}_{ij} \enspace .
\end{eqnarray}
We try to compute~(\ref{eqn:ZCmatrix}) community by community. So, we obtain
\begin{eqnarray*}
  F_{\textsc{zc}}(X) & = & \sum_{i,j \in V} \big( w_{ij}x_{ij} + (\mathcal{W}-w_{ij})(1-x_{ij}) \big) \\
  & = & \sum_{i,j \in V} ( 2 w_{ij}x_{ij} + \mathcal{W} - \mathcal{W} x_{ij} - w_{ij} ) \\
  & = & \sum_{i,j \in V} \big( (2 w_{ij} - \mathcal{W}) x_{ij} + \mathcal{W} - w_{ij} \big) \\
  & = & \sum_{i,j \in V} (2 w_{ij} - \mathcal{W}) x_{ij} + \mathcal{W} \cdot n^2 - 2m \enspace ,
\end{eqnarray*}
where $2m=\sum_{i,j \in V}w_{ij}$ for a weighted graph.

Now, given a partition $\mathcal{P}$ in more than two communities, $\sum_{i,j \in V} w_{ij} x_{ij}$ can be turned to $\sum_{C \in \mathcal{P}} \inC$ and $\sum_{i,j \in V} x_{ij}$ can be turned to $\sum_{C \in \mathcal{P}} \wC^2$. Thus, we obtain
\begin{eqnarray}
  \label{eqn:ZCcomms}
  F_{\textsc{zc}}(\mathcal{P}) & = & \sum_{C \in \mathcal{P}} (2 \cdot \inC - \mathcal{W} \cdot \wC^2) + \mathcal{W} \cdot n^2 - 2m \enspace .
\end{eqnarray}

Finally, we can easily deduce from~(\ref{eqn:ZCcomms}) the gain computation formula given in Algorithm~\ref{algo:ZCfunctions}, by reducing the following equation:
\begin{equation}
  \label{eqn:ZCgain}
  \left[ 2(\inC + 2 \cdot \dnc) - \mathcal{W}(\wC+s_i)^2 \right] - \left[ 2 \cdot \inC - \mathcal{W} \cdot \wC^2 - \mathcal{W} \cdot s_i^2 \right] \enspace .
\end{equation}
     
\medskip

Clearly, for the \textsl{Zahn-Condorcet} quality function, all the other communities $C^{\prime} \in (\mathcal{P} \setminus \{C\})$ are not impacted by the addition of $i$ to $C$ (i.e. their vectors \textsf{in} and \textsf{s} are not modified). Indeed, we can compute the gain to add any node $i$ to any community $C$ locally, i.e. without impacting the other communities. This is due to the fact that this criterion is linear, which is not the case for all the criteria.

We now show that all linear criteria can be integrated efficiently to Louvain.

\subsection{A Sufficient Condition to Integrate Efficiently any Criterion to Louvain}
\label{sec:suffcond}

The computation power of Louvain algorithm is clearly based on the fact that, at each step, we only compute a local quality, depending only on one node and one community: the Louvain method would not be so powerful if we had to compute the global quality of the partition at each step. Hence, for each criterion we want to integrate to Louvain, we have to express it locally, i.e. only for one node and one community. But how (and when) can we do that?

Let $F$ be any quality function. As we have explained in Sect.~\ref{sect:criteria}, this function depends on the weighted adjacency matrix of the graph $\textbf{W}$ and the unknown variable $\textbf{X}$. If $F$ is a linear function of $\textbf{X}$, then, $F$ can be computed by summing products for each $(i,j) \in V \times V$. Therefore, we are able to compute $F$ only for one part of $\textbf{W}$ and $\textbf{X}$, that can easily correspond to one community.

So, any quality function that verifies \emph{linearity} can be implemented efficiently in the Louvain framework.

\medskip

This condition is sufficient, but not necessary. Indeed, several quality functions are not linear but separable (i.e. they can be written as a scalar product of a part of $\phi(\textbf{W})$ and $\psi(\textbf{X})$): the fact that they can be efficiently plugged to Louvain strongly depends on the functions $\phi$ and $\psi$ applied on $\textbf{W}$ and $\textbf{X}$.

We give in Sect.~\ref{sec:counter} an example of a non-linear (but separable) criterion which is efficiently pluggable to Louvain.

\subsection{Several non-linear criteria (but separable) can be efficiently integrated to Louvain}
\label{sec:counter}

Algorithm~\ref{algo:PDfunctions} gives the four functions (INIT, REMOVE, INSERT and GAIN) designed for the Profile Difference criterion (\inCp and \dncp correspond to the \inC and \dnc modified values: see explanations below).

\begin{algorithm}[t]

  \caption{Functions for the Profile Difference quality function}
  \label{algo:PDfunctions}
      
  \DontPrintSemicolon
      
  INIT$(i)$:\;
  \hspace*{1mm}\Begin{
      $\textsf{in}^{\prime}[i] \leftarrow \hat{w}_{ii}$\;
      $\textsf{s}[i] \leftarrow s_i$\;
      $\kappa = n$\;
    }
    
    \BlankLine

  REMOVE$(i,C)$:\;
  \hspace*{1mm}\Begin{
      $\inCp \leftarrow \inCp - \dncp$\;
      $\wC \leftarrow \wC - s_i$\;
      \If{$\wC = 0$}{
        $\kappa \leftarrow \kappa - 1$
      }
      $\mathcal{P}[i] \leftarrow \emptyset$\;
    }
    
    \BlankLine
    
    INSERT$(i,C)$:\;
    \hspace*{1mm}\Begin{
        $\inCp \leftarrow \inCp + \dncp$\;
        \If{$\wC = 0$}{
          $\kappa \leftarrow \kappa + 1$
        }
        $\wC \leftarrow \wC + s_i$\;
        $\mathcal{P}[i] \leftarrow C$\;
      }
      
      \BlankLine
      
      GAIN$(i,C)$:\;
      \hspace*{1mm}\Begin{
          \eIf{$\wC = 0$}{
            return~$\displaystyle \frac{\dncp}{s_i} - \frac{1}{2}$\;
          }
          {
            return~$\displaystyle \frac{\inCp + \dncp}{\wC + s_i} - \frac{\inCp}{\wC}$\;
          }
        }
        
      \end{algorithm}

We recall here (for the weighted case) the Relational notation of the Profile Difference criterion:
\begin{eqnarray}
  \label{eqn:PDmatrix}
  F_{\textsc{pd}}(X) & = & 2 \sum_{i,j \in V} \frac{\hat{w}_{ij} x_{ij}}{|C_{[i]}|} - \kappa - \sum_{i,j \in V} \hat{w}_{ij}^2 \enspace ,
\end{eqnarray}
with $\kappa = |\mathcal{P}|$ and $C_{[i]}$ the community of node $i$.

Here, the $\phi$ (resp. $\psi$) function corresponds to the transformation of $\textbf{W}$ (resp. $\textbf{X}$) to $\hat{\textbf{W}}$ (resp. $\hat{\textbf{X}}$). In other words,
\begin{equation*}
\phi_{\textsc{pd}}(w_{ij})=\hat{w}_{ij}~=~\frac{2 w_{ij}}{d_i + d_j} \qquad \text{and} \qquad \psi_{\textsc{pd}}(x_{ij})=\hat{x}_{ij}~=~\frac{x_{ij}}{|C_{[i]}|} \enspace .
\end{equation*}
The $\phi_{\textsc{pd}}$ function can clearly be done as a pretreatment of Louvain (precisely before running Algorithm~\ref{algo:louvain}), since weights on links in the initial graph do not change during computation. We can also compute at the same time the constant $\mathcal{S} = \sum_{i,j \in V} \hat{w}_{ij}^2$ to further simplify computations. Moreover, the $\phi_{\textsc{pd}}$ and the $\psi_{\textsc{pd}}$ functions only apply local transformations. Therefore, we obtain
\begin{eqnarray}
  \label{eqn:PDcomms}
  F_{\textsc{pd}}(X) & = & 2 \sum_{C \in \mathcal{P}} \frac{\inCp}{\wC} - \kappa - \mathcal{S} \enspace ,
\end{eqnarray}
where \inCp denotes the \inC values related to the matrix $\hat{\textbf{W}}$ and where, in Algorithm~\ref{algo:PDfunctions},
\begin{eqnarray}
  \label{eqn:dncp}
  \dncp & = & 2 \sum_{j \in C} \hat{w}_{ij} + \hat{w}_{ii} \enspace .
\end{eqnarray}

Finally, we can easily deduce from~(\ref{eqn:PDcomms}) the gain computation formula given in Algorithm~\ref{algo:PDfunctions}.

\bigskip

Unfortunately, all the separable criteria cannot be integrated efficiently to the Louvain method. This is for example the case for the \textsl{Mancoridis-Gansner} quality function.

We recall here (for the weighted case) the Relational notation of the \textsl{Mancoridis-Gansner} criterion:
\begin{eqnarray}
  \label{eqn:MGmatrix}
  F_{\textsc{mg}}(X) & = & \frac{1}{\kappa}\sum_{i,j \in V} \frac{w_{ij}x_{ij}}{|C_{[i]}| \cdot |C_{[j]}|}+\frac{1}{\kappa(\kappa-1)} \sum_{i,j \in V} \frac{\bar{w}_{ij}\bar{x}_{ij}}{|C_{[i]}| \cdot |C_{[j]}|} \enspace ,
\end{eqnarray}
with $\kappa > 1$ the number of communities in the partition.

Here, $\phi(w_{ij}) = w_{ij}$ and $\bar{\phi}(\bar{w}_{ij}) = \bar{w}_{ij}$. However, the $\psi$ and the $\bar{\psi}$ functions turn $x_{ij}$ (resp. $\bar{x}_{ij}$) to $\frac{x_{ij}}{|C_{[i]}| \cdot |C_{[j]}|}$ (resp. $\frac{\bar{x}_{ij}}{|C_{[i]}| \cdot |C_{[j]}|}$). We can simplify $\frac{x_{ij}}{|C_{[i]}| \cdot |C_{[j]}|}$ as $\frac{x_{ij}}{|C_{[i]}|^2}$, since the nodes $i$ and $j$ are here in the same community (thus, $C_{[i]} = C_{[j]}$). Hence, we obtain
\begin{eqnarray*}
  F_{\textsc{mg}}(X) & = & \frac{1}{\kappa}\sum_{i,j \in V} \frac{w_{ij}x_{ij}}{|C_{[i]}|^2}+\frac{1}{\kappa(\kappa-1)} \sum_{i,j \in V} \frac{\bar{w}_{ij}\bar{x}_{ij}}{|C_{[i]}| \cdot |C_{[j]}|} \\
  & = & \frac{1}{\kappa}\sum_{i,j \in V} \frac{w_{ij}x_{ij}}{|C_{[i]}|^2} + \frac{1}{\kappa(\kappa-1)} \sum_{i,j \in V} \frac{(\mathcal{W}-w_{ij})(1-x_{ij})}{|C_{[i]}| \cdot |C_{[j]}|} \\
  & = & \frac{1}{\kappa}\sum_{i,j \in V} \frac{w_{ij}x_{ij}}{|C_{[i]}|^2} + \frac{1}{\kappa(\kappa-1)} \sum_{i,j \in V} \left( \frac{(w_{ij} - \mathcal{W})x_{ij}}{|C_{[i]}|^2} + \frac{\mathcal{W} - w_{ij}}{|C_{[i]}| \cdot |C_{[j]}|} \right) \enspace .
\end{eqnarray*}

Now, $\sum_{i,j \in V} w_{ij} x_{ij}$ can be turned to $\sum_{C \in \mathcal{P}} \inC$ and $\sum_{i,j \in V} x_{ij}$ can be turned to $\sum_{C \in \mathcal{P}} \wC^2$. Moreover, $|C_{[i]}|^2$ can be turned to $\wC^2$ again. Thus, we have
\begin{eqnarray*}
  F_{\textsc{mg}}(X) & = & \frac{1}{\kappa}\sum_{C \in \mathcal{P}} \frac{\inC}{\wC^2} + \frac{1}{\kappa(\kappa-1)} \left( \sum_{C \in \mathcal{P}} \frac{\inC - \mathcal{W} \cdot \wC^2}{\wC^2} + \sum_{i,j \in V} \frac{\mathcal{W} - w_{ij}}{|C_{[i]}| \cdot |C_{[j]}|} \right) \\
  & = & \frac{1}{\kappa}\sum_{C \in \mathcal{P}} \frac{\inC}{\wC^2} + \frac{1}{\kappa(\kappa-1)} \left( \sum_{C \in \mathcal{P}} \frac{\inC}{\wC^2} - \mathcal{W} \cdot \kappa + \mathcal{W} \cdot \kappa^2 - \sum_{i,j \in V} \frac{w_{ij}}{|C_{[i]}| \cdot |C_{[j]}|} \right) \\
  & = & \frac{1}{\kappa}\sum_{C \in \mathcal{P}} \frac{\inC}{\wC^2} + \frac{1}{\kappa(\kappa-1)} \left( \sum_{C \in \mathcal{P}} \frac{\inC}{\wC^2} - \sum_{i,j \in V} \frac{w_{ij}}{|C_{[i]}| \cdot |C_{[j]}|} \right) + \mathcal{W} \enspace .
\end{eqnarray*}
The term $\sum_{i,j \in V} \frac{w_{ij}}{|C_{[i]}| \cdot |C_{[j]}|}$ cannot be simplified: we are not able to group by summing here the $w_{ij}$ terms community by community, since inter-cluster links are also considered (i.e. for several nodes $i$ and $j$, we have $|C_{[i]}| \ne |C_{[j]}|$). That implies for each modification, we have to compute the \textsl{Mancoridis-Gansner} quality function for the whole graph, and not only for one community, since a local change may impact many communities.

\medskip

Moreover, several criteria are restricted to non-weighted graphs (see Table~\ref{tabCriterLinSep} page~\pageref{tabCriterLinSep}), or need to fix the number of communities in advance, like the \textsl{Michalski-Decaestecker} criterion (see Table~\ref{tabCriterSepNonLin} page~\pageref{tabCriterSepNonLin}), that is no possible with the Louvain method (report to the Sect.~\ref{sect:criteria} for more details).

\bigskip

To summary, the adaptation of criteria in terms of practicity is subject to one of the following conditions:
\begin{enumerate}
\item linearity,
\item separability, but the $\psi$ function depends only upon the community in study, in terms of links, size or degree.
\end{enumerate}

%% file: exp.tex
\section{Experimentations}
\label{sect:exp}

In this section, we present and analyze results of experiments made on a number of testcase networks.

\subsection{Description and Technical Characteristics}

We have run the generic Louvain method on the same networks as in~\cite{ABGL-2011,BGLL-2008} (see Table~\ref{tab:instances} for details) and we have used the following criteria (report to Sect.~\ref{sect:criteria} page~\pageref{sect:criteria} for a more complete description):
\begin{enumerate}
\item the classical \textsl{Newman-Girvan} modularity (denoted in the following by \textsc{ng}),
\item the \textsl{Zahn-Condorcet} criterion (denoted by \textsc{zc}),
\item the Deviation to Indetermination quality function (denoted by \textsc{di}),
\item the Deviation to Uniformity criterion (denoted by \textsc{du}),
\item the Balanced Modularity (denoted by \textsc{bm}),
\item the \textsl{Michalski-Goldberg} criterion (denoted by \textsc{g}),
\item the Profile Difference quality function (denoted by \textsc{pd}).
\end{enumerate}

The software generic Louvain method is available at~\cite{gen-louvain}. As for the classical Louvain algorithm, they are written in C++ language, using classical STL containers.

The computer used for the experimentations has a Quad Core processor running at $\np{2.93}$ GHz, $4$ MB cache memory and $60$ GB RAM.

\begin{table}[!ht]
  \caption{Properties of instances used for experimentations}

  \begin{center}
    \begin{tabular}{|c||c|c|c|}
      \hline
      & $n$ & $m$ & Memory space used for storage$^a$\\\hline\hline
      \texttt{karate}~\cite{zachary1977information} & $34$ & $78$ & \np{3,48} KB\\\hline
      \texttt{arxiv}~\cite{arxiv} & $\np{9377}$ & $\np{48214}$ & \np{1.95} MB\\\hline
      \texttt{internet}~\cite{HM-2003} & $\np{69949}$ & $\np{351380}$ & \np{14,20} MB\\\hline
      \texttt{webndu}~\cite{AJB-1999} & $\np{325730}$ & $\np{2180216}$ & \np{86.90} MB \\\hline\hline
      \texttt{webuk-2005}~\cite{webuk-2005} & $\np{39459925}$ & $\np{783027125}$ & \np{29.61} GB \\\hline
      \texttt{webbase-2001}~\cite{webbase-2001} & $\np{118142155}$ & $\np{1019903190}$ & \np{39.31} GB\\\hline
    \end{tabular}
  \end{center}
  
  $^a${\footnotesize For each graph, we must store in the generic Louvain framework:
    \begin{itemize}
    \item degree of each node, using \texttt{unsigned long long} data type,
    \item neighbors (labels and links weight) of each node, using respectively \texttt{int} and \texttt{long double} data types,
    \item weight of each node, using \texttt{int} data type.
    \end{itemize}
    Moreover, we store the number of nodes $n$ (resp. links $2m$), using \texttt{int} (resp. \texttt{unsigned long long}) data type; we store also the sum of nodes (resp. links) weights, using \texttt{int} (resp. \texttt{long double}) data type. \\
    Thus, each graph takes $(2m+1)(4+16)+(n+1)(4+8) = 40m+12n+32$ bytes in memory.
  }
  
  \label{tab:instances}
\end{table}

\subsection{Results and Discussion}

Tables~\ref{tab:exp_times} and~\ref{tab:exp_kappa} summarize respectively the total (i.e. real) running times and the number of communities obtained for each quality function. We have run Louvain with a value of $5 \cdot 10^{-3}$ for the quality precision parameter (the default value is $10^{-6}$). Moreover, for the \texttt{karate}, \texttt{arxiv}, \texttt{internet} and \texttt{webndu} (resp. \texttt{webuk-2005} and \texttt{webbase-2001}) instances, values reported to Tables~\ref{tab:exp_times} and~\ref{tab:exp_kappa} corresponds to an average computed after ten (resp. three) executions.

\begin{table}[!ht]
  \centering
  \begin{tabular}{|c||c|c|c|c|c|c||c|}
    \hline
    & \textsc{ng} & \textsc{zc} & \textsc{di} & \textsc{du} & \textsc{bm} & \textsc{g} & \textsc{pd} \\\hline\hline
    \texttt{karate} & 0 & 0 & 0 & 0 & 0 & 0 & 0 \\\hline
    \texttt{arxiv} & 0 & 0 & 0 & 0 & 0 & 0 & \np{0.5} \\\hline
    \texttt{internet} & 0 & \np{0,5} & 0 & 0 & \np{0,5} & 2 & 1 \\\hline
    \texttt{webndu} & 1 & 2 & 1 & \np{1} & 1 & \np{2.5} & \np{2.8} \\\hline\hline
    \texttt{webuk-2005} & \np{232} & \np{468} & \np{201} & \np{192} & \np{178} & \np{171439} & \np{153174} \\\hline
    \texttt{webbase-2001} & \np{430} & \np{1043} & \np{446} & \np{394} & \np{324} & \np{5736} & -- \\\hline
  \end{tabular}
  \caption{Running times (in seconds) obtained for each criterion}
  \label{tab:exp_times}
\end{table}

\begin{table}[!ht]
  \centering
  \begin{tabular}{|c||c|c|c|c|c|c||c|}
    \hline
    & \textsc{ng} & \textsc{zc} & \textsc{di} & \textsc{du} & \textsc{bm} & \textsc{g} & \textsc{pd} \\\hline\hline
    \texttt{karate} & \np{4} & \np{19} & \np{4} & \np{5} & \np{4} & \np{10} & \np{3} \\\hline
    \texttt{arxiv} & \np{58} & \np{2557} & \np{64} & \np{77} & \np{54} & \np{3301} & \np{937} \\\hline
    \texttt{internet} & \np{47} & \np{40619} & \np{43} & \np{179} & \np{38} & \np{21204} & \np{2917} \\\hline
    \texttt{webndu} & \np{450} & \np{201728} & \np{329} & \np{759} & \np{318} & \np{53273} & \np{5205} \\\hline\hline
    \texttt{webuk-2005} & \np{19844} & \np{21738671} & \np{80850} & \np{72838} & \np{21355} & \np{7044755} & \np{545223} \\\hline
    \texttt{webbase-2001} & \np{2762172} & \np{71807969} & \np{2742902} & \np{2786272} & \np{2737359} & \np{40507623} & -- \\\hline
  \end{tabular}
  \caption{Number of community obtained for each criterion}
  \label{tab:exp_kappa}
\end{table}

\medskip

The Profile Difference criterion is isolated since it needs a pretreatment on the initial graph to work. This operation can be very long on large instances. Moreover, it requires a lot of memory space, that explains why it fails on the \texttt{webbase-2001} instance.

\bigskip

Except for the \textsl{Michalski-Goldberg} criterion (and the other criterion mentioned above), all the quality functions tested here present reasonnable running times. Indeed, they do not exceed twenty minutes on the biggest instance (which has, however, more than $100$ millions nodes and $1$ billion links). Moreover, they are close to the classical modularity running times; the Balanced Modularity and the Deviation to Uniformity criterion are even faster than it. Hence, the generic Louvain method presented in this paper is quite suitable to find communities in large complex networks.

\medskip

The difference between running times could be related to the number of communities obtained. Indeed, \textsc{zc} and \textsc{g}, which are not subject to the resolution limit problem, present bigger running times than the others. However, on the \texttt{webuk-2005} instance, we can observe some differences:
\begin{enumerate}
\item \textsc{bm} is faster than \textsc{ng}, but they produce similar number of communities;
\item \textsc{di} and \textsc{du} present similar running times than \textsc{ng} and \textsc{bm}, but they generate more communities.
\end{enumerate}

Moreover, \textsc{zc} and \textsc{g} generate a lot of communities, but the first one is clearly faster. Actually, we think that differences between running times observed here could be due to the intrinsic design of the quality functions. More precisely, we think that criteria which optimize both the number of intra-cluster links and the number of inter-cluster missing links can converge faster than the others. This is anyway the case for \textsc{zc} and \textsc{bm} (conversely, \textsc{g} optimizes only the intra-cluster density and converges slower than the other). But a more complete study focused on criteria is needed to validate (or not) this fact, and more generally to try to understand the observed phenomena.

%% file: conclusion.tex
\section{Conclusion and Future Works}
\label{sect:conclusion}

We have presented a new version of the Louvain method, which can approach the optimal solution of many quality functions for graph partitioning. For that, we have shown that the Louvain method core is mainly independent of the underlying quality function, and therefore that it can be generalized to many criteria. Moreover, we have given a sufficient condition which ensures that a quality function can be efficiently plugged to the Louvain framework. We have also given examples of quality function implementations, including a criterion which does not meet the sufficient condition. Then, we have shown that the generic Louvain method is very efficient, since for most criteria, its running times do not exceed twenty minutes, even on huge instances with more than one billion links.

\medskip

This new version of the Louvain method offers multiple benefits. As we are able to optimize many quality functions (not only the classical modularity) on huge graphs, this allows us to find different kind of communities (and, therefore, this can improve partitions quality); for instance, it allows to find good partitions on specific complex networks on which the \textsl{Newman-Girvan} modularity fails. More generally, the scope of the generic Louvain method is larger; it is worth recalling that the criteria list presented in this paper is not exhaustive: many other quality functions can be integrated to the Louvain framework.

\bigskip

In Sect.~\ref{sect:exp}, we have focused on running times and size partitions. We plan to make a more complete study focused on criteria. For that, it would be interesting to compare them on complex networks with known communities, and also on random graphs with communities (e.g., using the \textsl{LFR} benchmark~\cite{LFR-2008}).

Experiments done on huge random graphs (using different models with different properties: \textsl{Erd\H{o}s-R\'{e}nyi}, configuration model, \textsl{Albert-Barab\'{a}si}, etc.) or theoretical graph families would also allow to gain in depth insight of the quality functions behavior: number of iterations and levels done, convergence time, etc.

\medskip

Eventually, our work can open the way to other forms of Louvain algorithm generalization, e.g., for dynamic complex networks.